\documentstyle[sprocl,epsfig]{article}

\def\Journal#1#2#3#4{{#1} {\bf #2}, #3 (#4)}

\def\PRD{{\em Phys. Rev.} D}

\def\be{\begin{equation}}
\def\ee{\end{equation}}
\def\bea{\begin{eqnarray}}
\def\eea{\end{eqnarray}}
\def\ttb{\hbox{$t {\bar t}$}}
\def\dlde{\hbox{d${\cal L}$/dE}}

\begin{document}

\title{The \ttb\ Threshold and Machine Parameters at the NLC}

\author{ D.Cinabro }

\address{Wayne State University, Detroit,\\ MI 48202, USA}


\maketitle\abstracts{
One of the problems in the design of a high energy $e^+e^-$ linear collider is
the distribution of the luminosity as a function
of real collision energy, \dlde, due to initial state
radiation, beamstrahlung, and the energy spread of
the collider.  These effects smear a sharp feature in the cross section
of $e^+e^- \to$ hadrons such as the \ttb\ threshold into a flatter
structure.  This study reviews the impact on \dlde\ of
these effects as a function of machine parameters,
explores some methods of measuring \dlde\ with Bhabha scattering,
how the d${\cal L}$/dE flattens the \ttb\ cross section near threshold,
and the effect the \dlde\ measurement has on 
extracting parameters of the top
quark, such as the mass and width, in a 2.5/fb scan of
the \ttb\ threshold.
}

%


	One of the prime goals of a high energy $e^+e^-$ linear collider
is the study of sharp features in the $e^+e^- \to$ hadrons cross
section.  The \ttb\ threshold is an excellent example of such a structure.
The cross section for $e^+e^- \to t \bar t$ is expected to rise by an order of
magnitude with only a 10 GeV change in center-of-mass energy around 350 GeV.
Careful study of this \ttb\ threshold structure can precisely measure many
parameters of the top quark, including its mass and width.\cite{aspen}

	One of the major problems in the design of a high energy $e^+e^-$
linear collider is the effects that cause the luminosity spectrum
of the collisions, \dlde, to be smeared.
One of these, initial state radiation, ISR, is simply the effect
of QED on the initial electron and positron in $e^+e^-$ collision and
cannot be controlled.  On the other hand beamstrahlung, electromagnetic
radiation by the particles in one bunch caused by their interaction with
the electric field of the bunch they are colliding with, and the energy
spread of the linac can be controlled by varying parameters of the collision.
These parameters, such as the size of collision region and the bunch current,
are also crucial in determining the luminosity of the machine.  In general
the parameters that lead to the highest luminosity also lead to largest
smearing of \dlde.\cite{param}

	This work studies the question of the impact of the machine
parameters of a high energy $e^+e^-$ linear collider on the extraction of
the parameters of a sharp feature in the $e^+e^- \to$ hadrons cross section.
First I review the effects that smear \dlde.  Next I
explore various techniques of measuring \dlde\ in Bhabha scattering.
Then I show the effect of \dlde\ on the \ttb\ threshold and the effect
the measurement of \dlde\ with Bhabha scattering has on the extraction of
top quark parameters in a scan of the \ttb\ threshold.
 

	There are three effects that smear the collision energy of any
$e^+e^-$ collider.  The first is a consequence of QED; the incoming electron
and positron can radiate photons leaving them at an energy below the nominal
beam energy.  This is called initial state radiation, ISR.
It is simulated using the Pandora\cite{pandora}
Monte Carlo which uses the Skrzypek-Jadach approximation for the energy
spectrum of the incoming beams.  ISR depends only on the nominal beam
energy and its effect on a 350 GeV center of mass collision is shown
in Figure 1(a).
  
\begin{figure}
\begin{picture}(100,175)(0,0)
\put(-10, 80){\epsfig{file=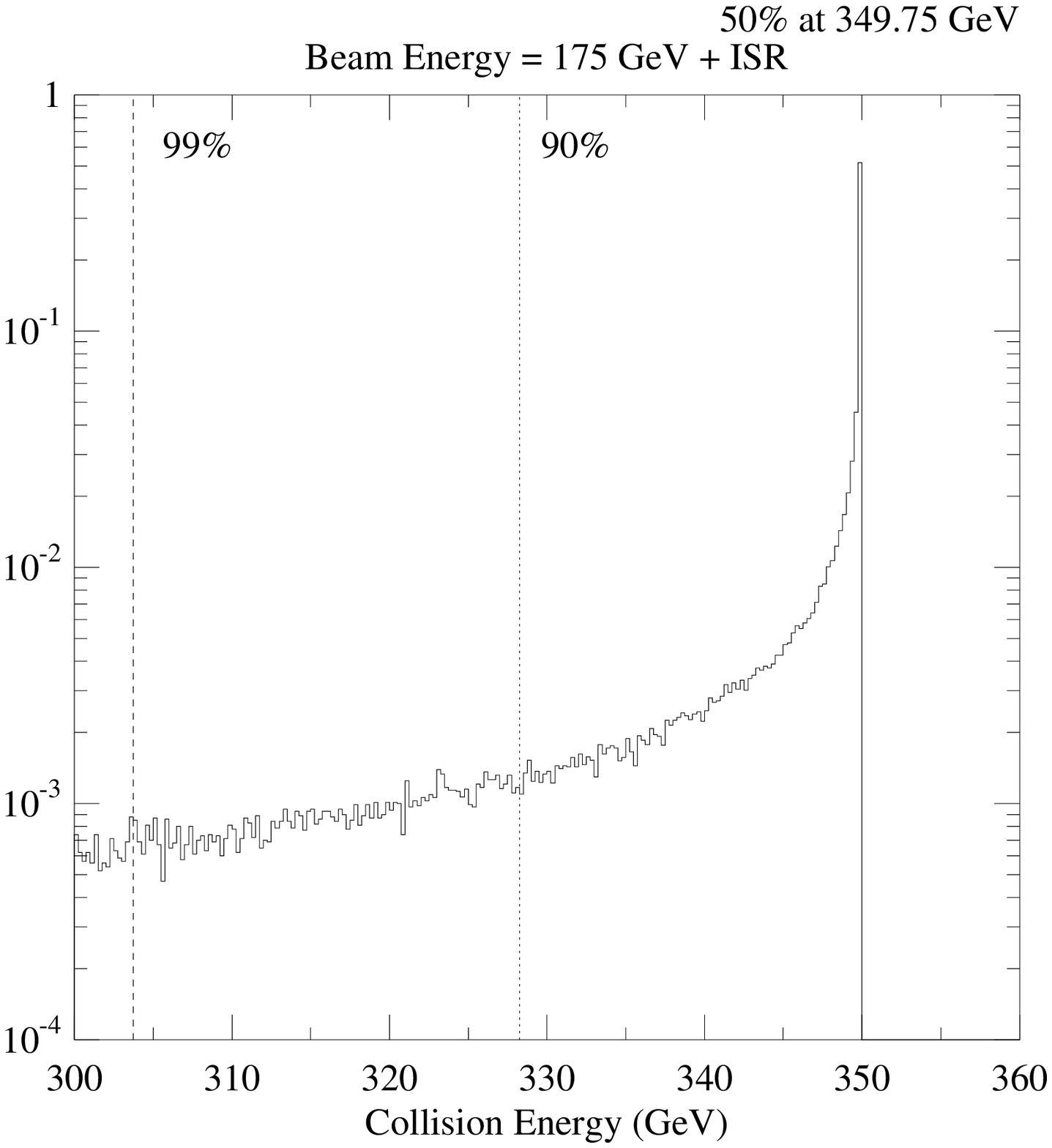,height=1.25in}}
\put (10,125){(a)}
\put(62.5, 80){\epsfig{file=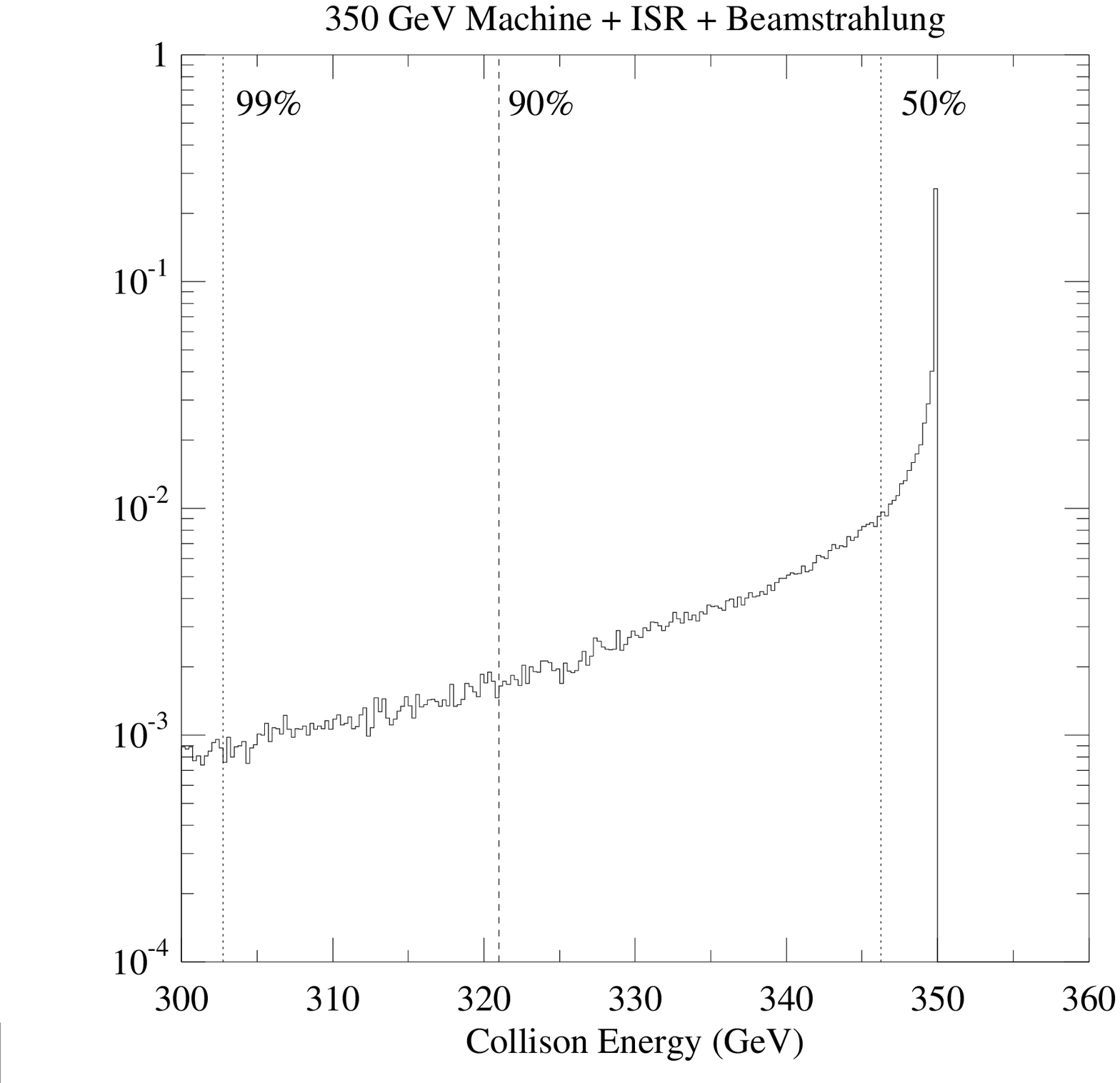,height=1.25in}}
\put(82.5,125){(b)}
\put(135, 80){\epsfig{file=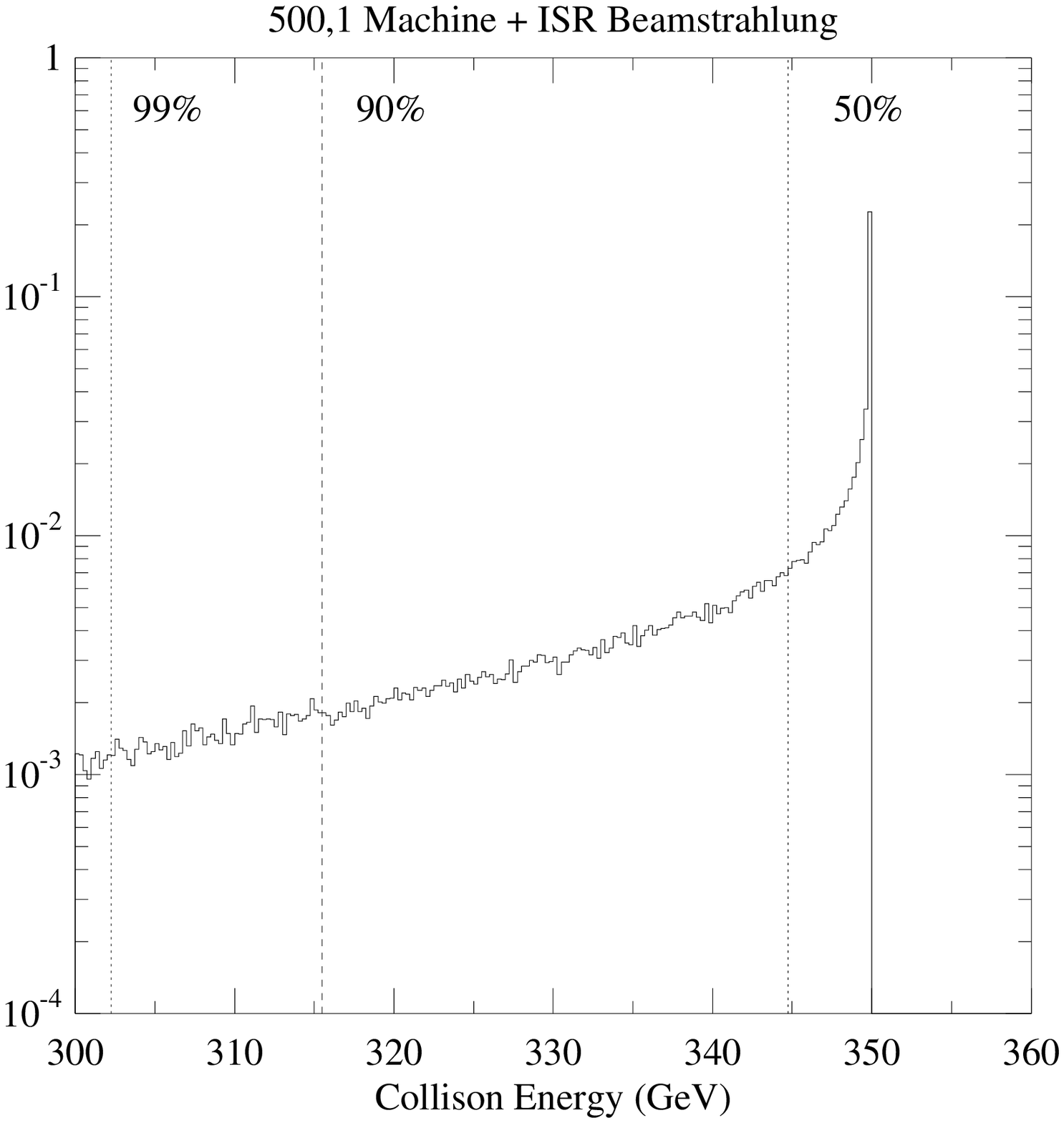,height=1.25in}}
\put(155,125){(c)}
\put(207.5, 80){\epsfig{file=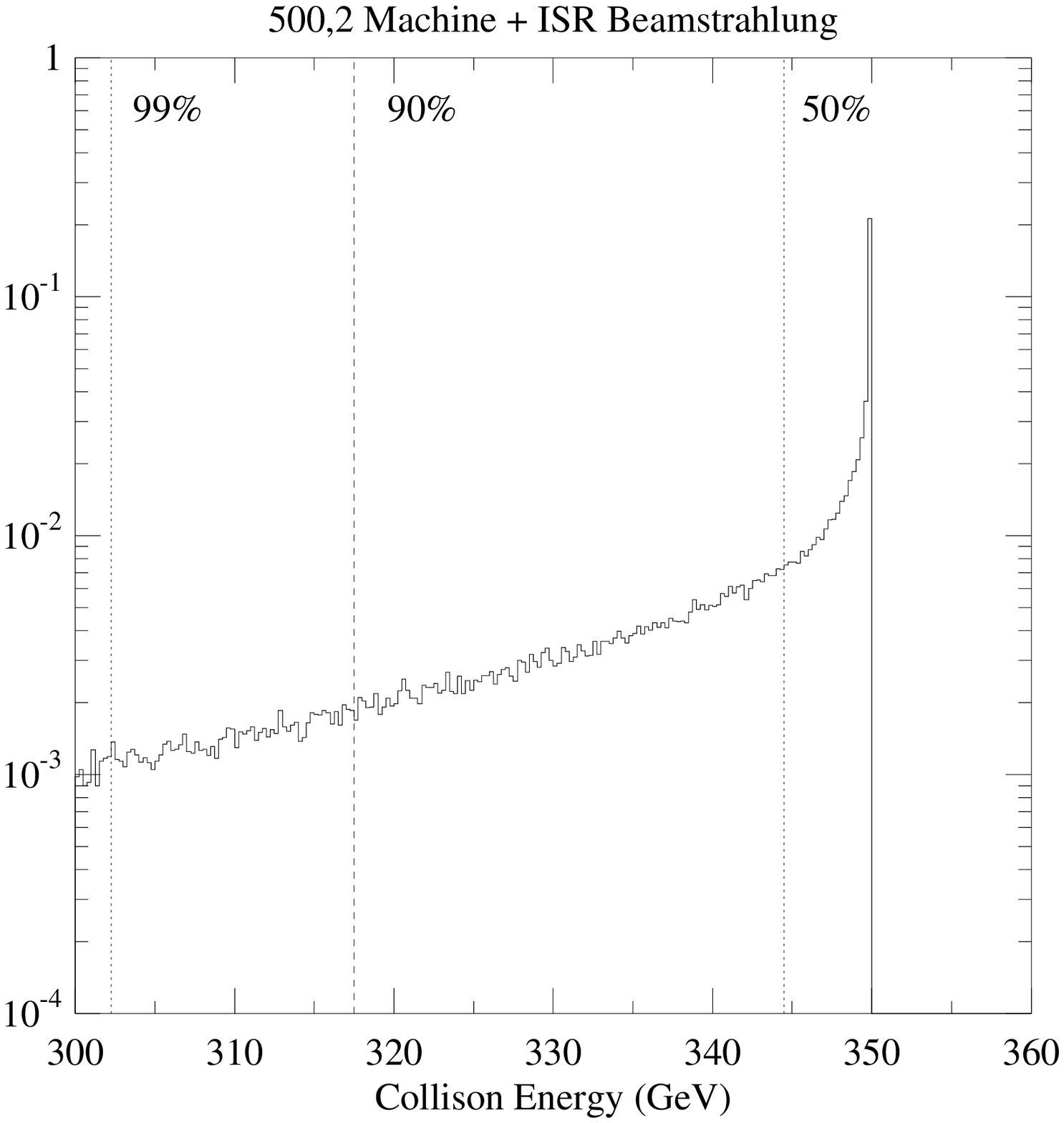,height=1.25in}}
\put(227.5,125){(d)}
\put(280,  80){\epsfig{file=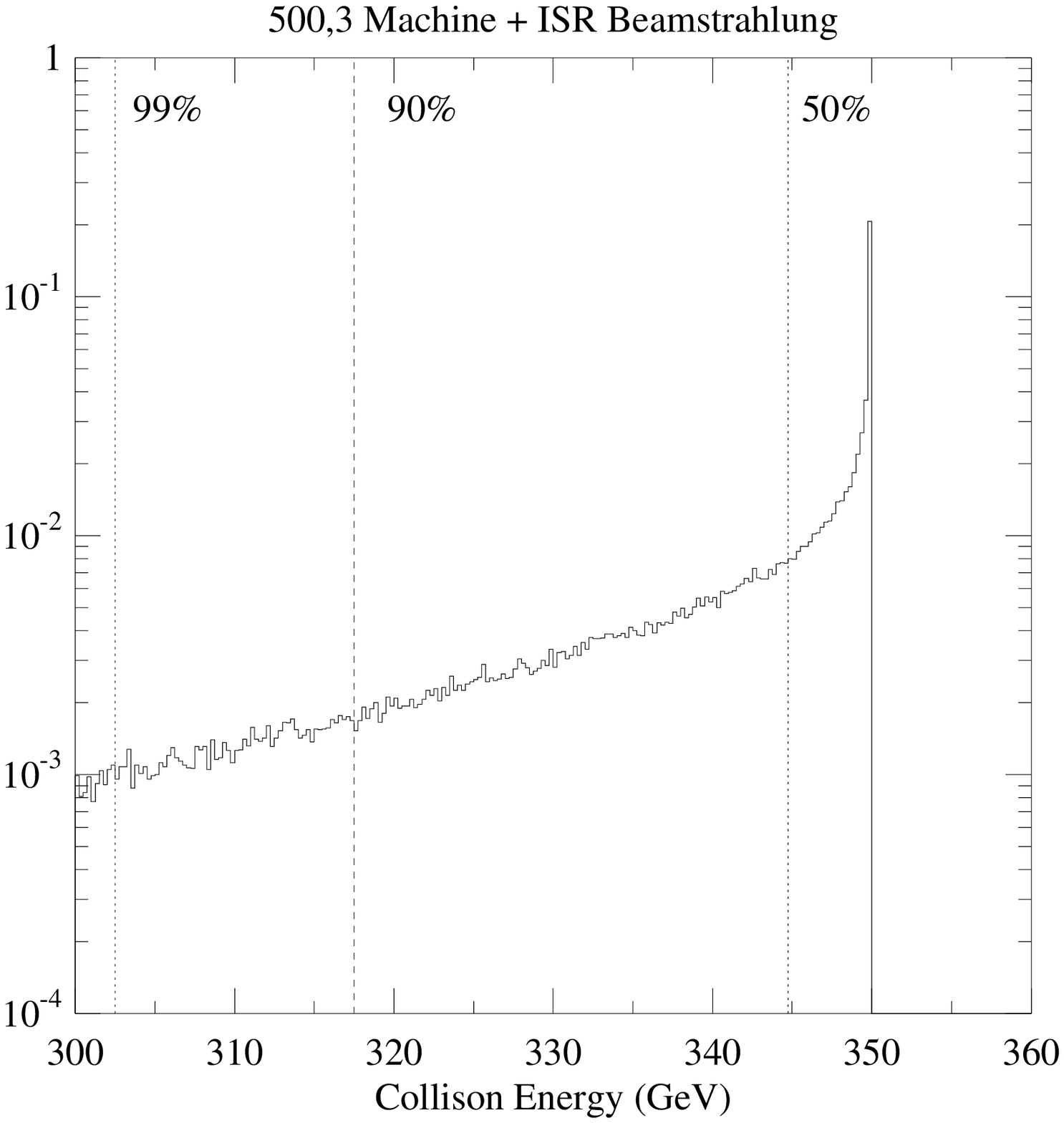,height=1.25in}}
\put(300,125){(e)}
\put( 20,  0){\epsfig{file=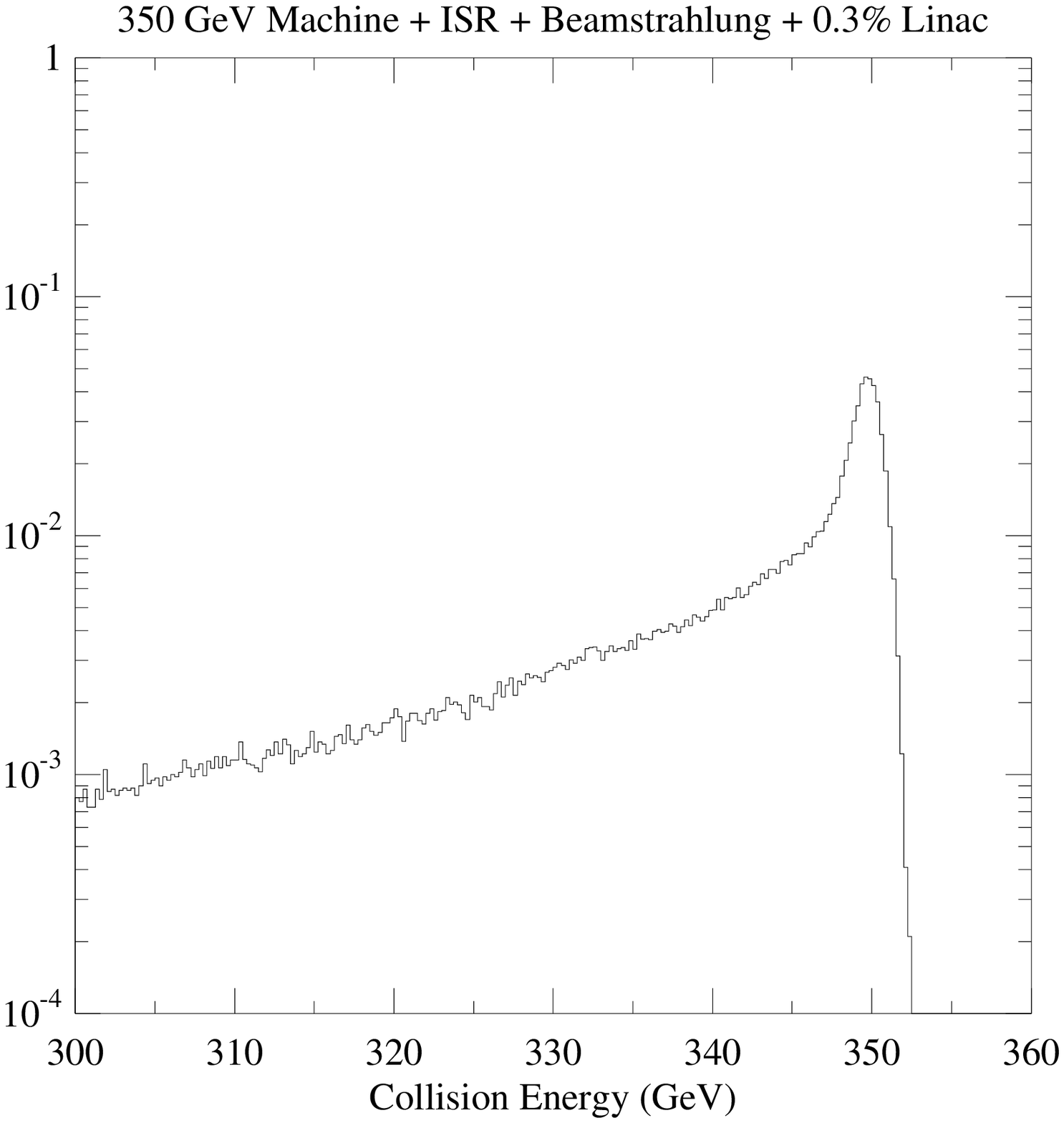,height=1.25in}}
\put( 40, 45){(f)}
\put(92.5, 0){\epsfig{file=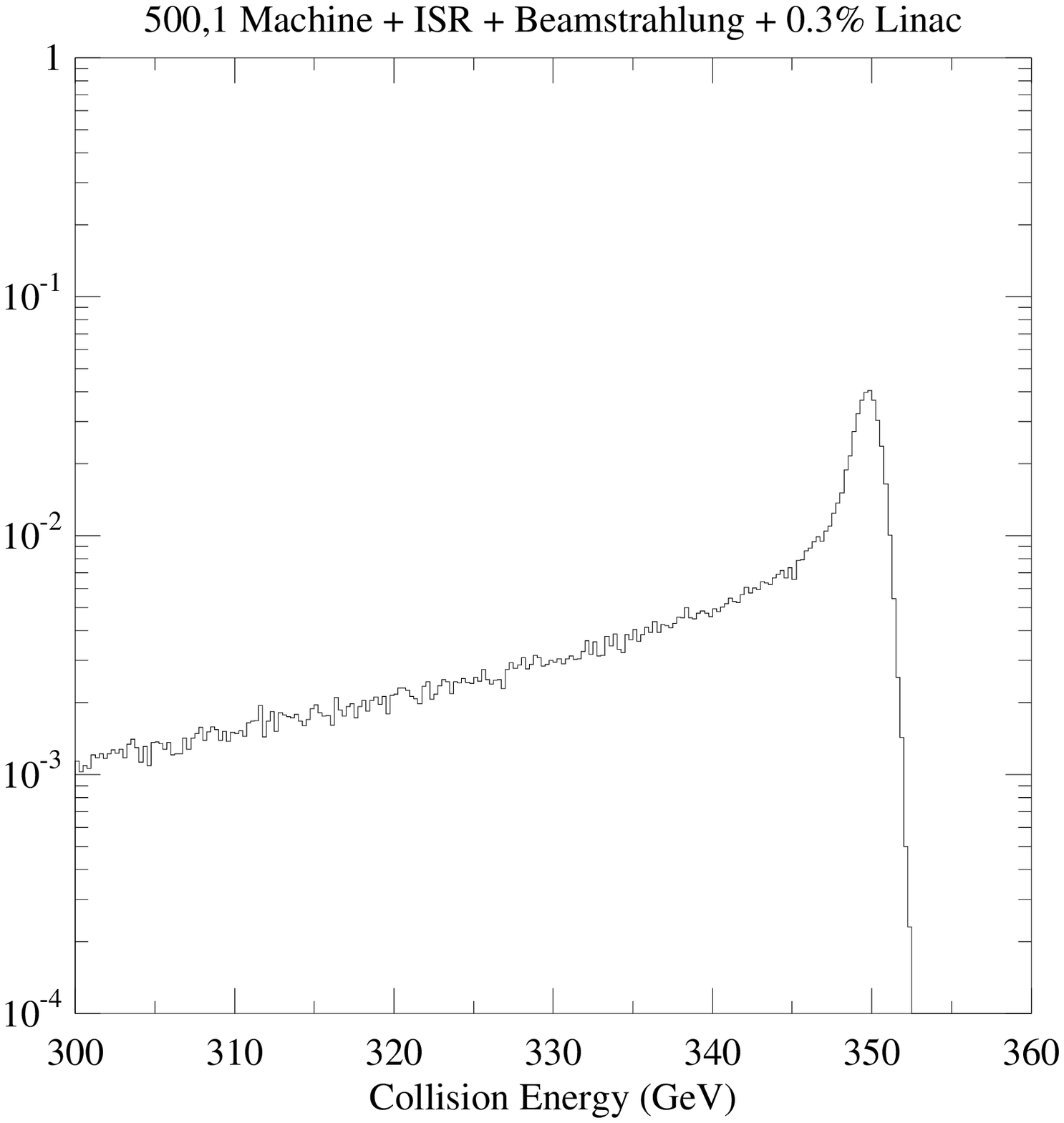,height=1.25in}}
\put(112.5,45){(g)}
\put(165,  0){\epsfig{file=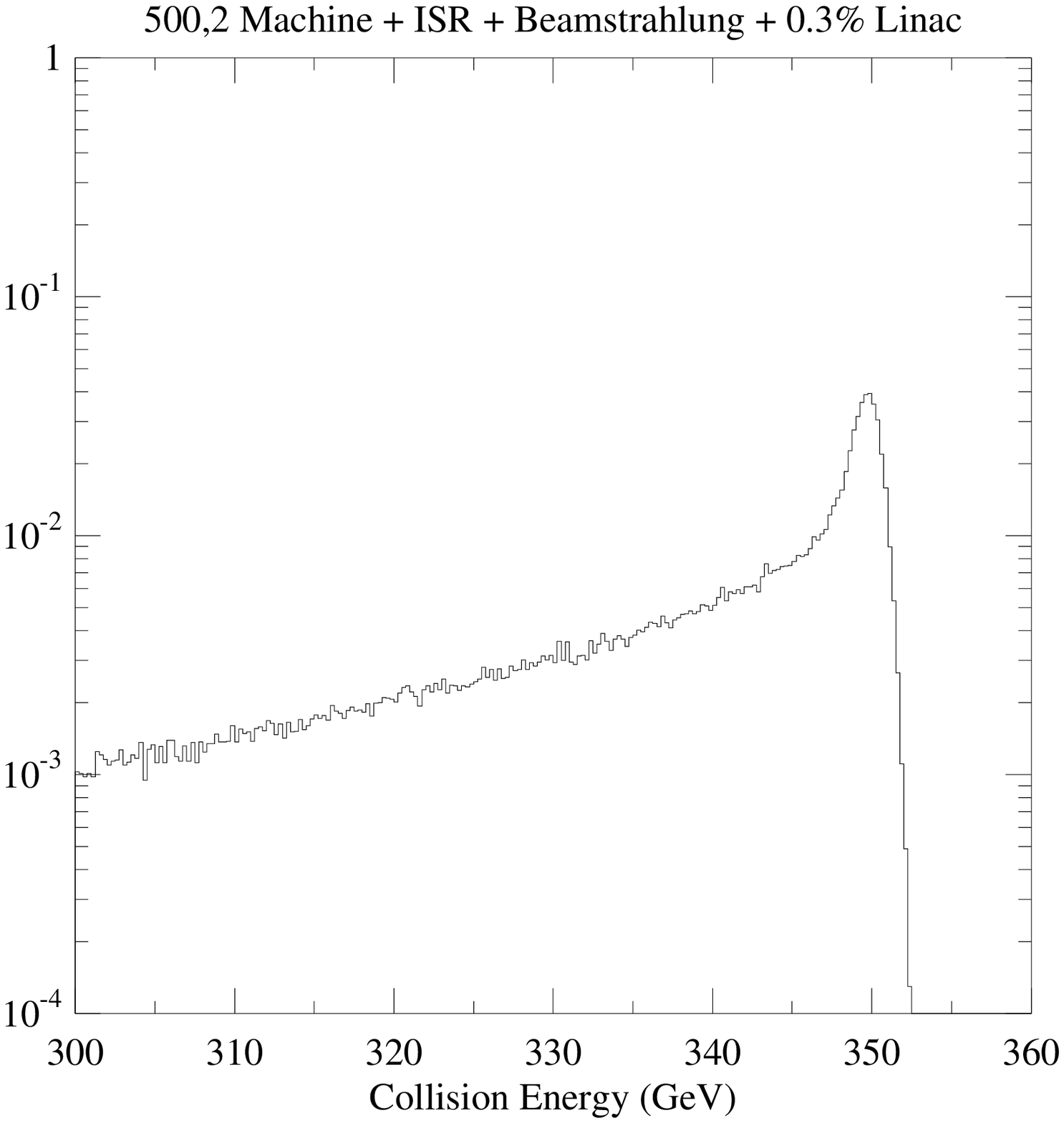,height=1.25in}}
\put(185, 45){(h)}
\put(237.5,  0){\epsfig{file=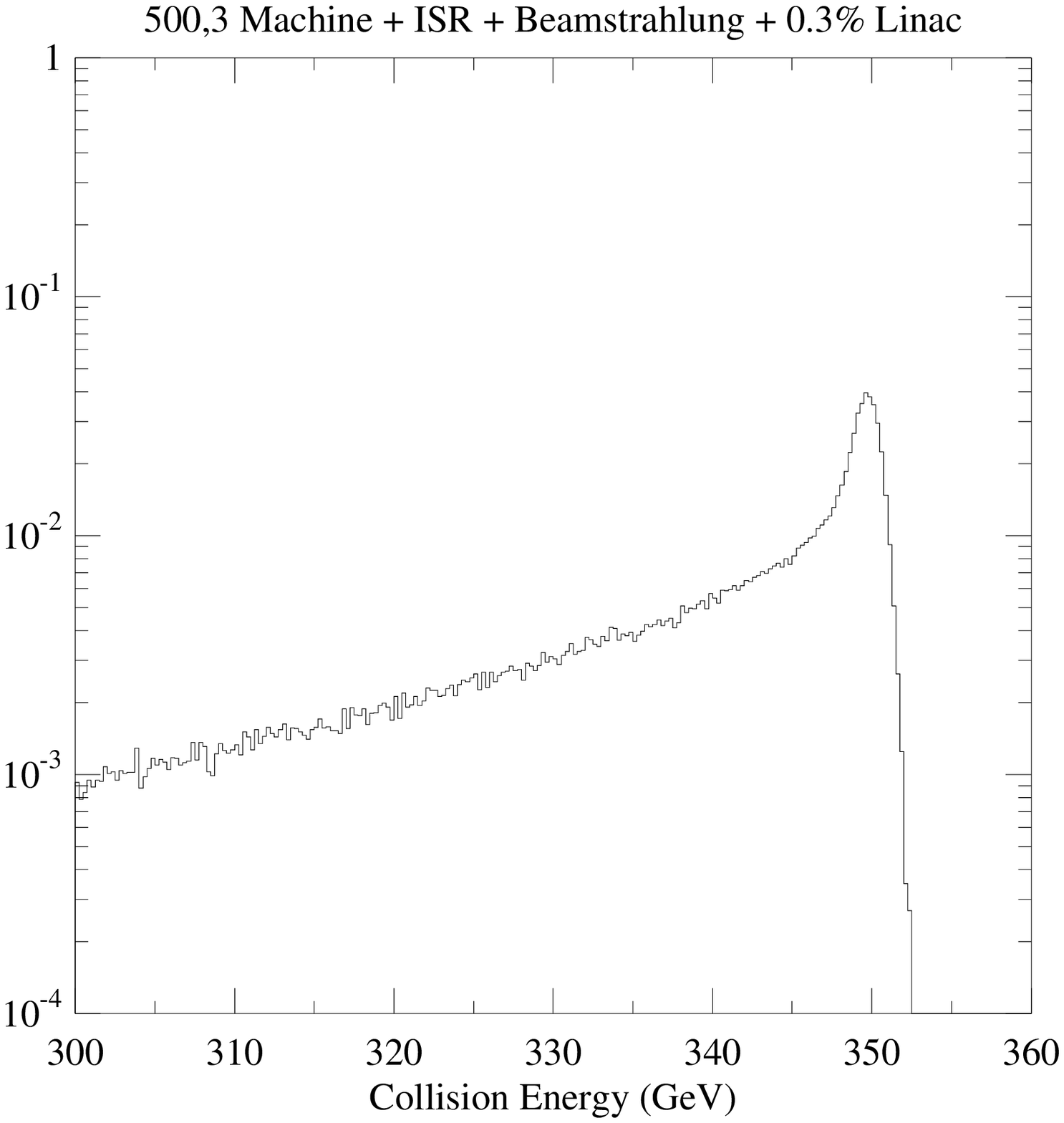,height=1.25in}}
\put(257.5,45){(i)}
\end{picture}
\caption{The \dlde\ spectrum for various effects and choices of
machine parameters.  (a) ISR only, (b-e) ISR and beamstrahlung for
parameter sets A-D respectively, (f-i) ISR, beamstrahlung and a
0.3\% linac energy spread for parameter sets A-D respectively.}
\end{figure}

	The second effect is the interaction of the particles in one beam
with the electric field generated by the other beam as the two collide.
This interaction causes the beams to radiate photons and is called
beamstrahlung.  Beamstrahlung is a large effect at a machine like the NLC
where the beam sizes are very small and thus produce very intense
fields.  It is controlled by the number of particles in the colliding
bunches, the relativistic factor gamma, and length and longitudinal sizes
of the beam.  These are used to calculate the disruption factor, $\Upsilon$,
the size of which controls the amount of beamstrahlung experienced by
the colliding beams.  Table~\ref{tab:beams} shows some possible
beam parameters for a 350 GeV nominal center of mass collision energy
NLC design.\cite{param}
\begin{table}[t]
\caption{Possible parameter choices for a 350 GeV NLC.\label{tab:beams}}
\vspace{0.4cm}
\begin{center}
\begin{tabular}{|c|c|c|c|c|}
\hline
Parameter Set        & A     & B     & C     & D     \\ \hline
N ($\times 10^{10}$) & 0.95  & 0.75  & 0.95  & 1.1   \\ 
$\sigma_x$ (nm)      & 402   & 246   & 327   & 365   \\ 
$\sigma_y$ (nm)      & 6.40  & 3.39  & 4.88  & 7.57  \\ 
$\sigma_z$ ($\mu$m)  & 120   & 90    & 120   & 145   \\ 
$\Upsilon$           & 0.061 & 0.094 & 0.075 & 0.065 \\
${\cal L} \times 10^{33}/{\rm cm}^2 {\rm sec}$
                     & 4.00  & 6.51  & 5.84  & 5.21 \\ \hline 
\end{tabular}
\end{center}
\end{table}
Note that in general as $\Upsilon$ and the size of the beamstrahlung
effect increases, the luminosity also increases.
Figures 1(b-e) show the effect on the \dlde\ spectrum of the
beamstrahlung effect on top of the ISR effect described above
respectively for the four possible NLC parameters given
in Table~\ref{tab:beams} using the Pandora Monte Carlo.

	The third effect at a high energy $e^+e^-$ linear collider is
the intrinsic energy spread of the linac.  An expectation for the
NLC is summarized in \cite{linac}.  In general the energy spread is not
Gaussian, but it can be accurately measured using synchrotron radiation,
and it is an effect is in the range of 0.1-0.5\% of the nominal beam energy.
For this study I simply smeared the incoming beam energy by a
Gaussian with an illustrative width of 0.3\% of the nominal
beam energy.  This effect on \dlde\ is shown in Figures~1(f-i) convoluted
with the effects of ISR and beamstrahlung respectively for
the four 350 GeV NLC parameter sets of Table~1. 


	Many methods have been proposed to extract the \dlde\ spectrum
at a high energy $e^+e^-$ linear collider.\cite{miller}  The
method which has been very effective at LEP is small angle Bhabha
scattering, $e^+e^- \to e^+e^-$, and I have explored using this
process at the NLC.  My method of exploration was to generate
about 1000 single 
electrons of 200 GeV and fully simulate their interactions in the
American S and L detectors\cite{SandL} in various polar angle
slices of the detector.  These events were then reconstructed using
JAS based reconstruction algorithms\cite{jas} and selections
of the single electron were made to choose well reconstructed
energies and momenta.  These selections resulted in a selection
efficiency and a resolution
on the energy of a single electron which depended on the polar
angle and the analysis technique.  These were then used
to smear the legs of $e^+e^- \to e^+e^-$ events generated with
the Pandora Monte Carlo.\cite{pandora}  Cross sections are also from
Pandora and are for both scattered particles to be in the given
polar angle range.  Four polar regions and three techniques are considered
in a 2.5/fb run at 350 GeV.

	The simplest analysis to consider is the barrel region.  For
example with $|\cos\theta| < 0.6$ the Bhabha cross section is
3.5 pb and the event energy is best measured with the tracking system.
Good single electrons
are selected by requiring one and only one track pointing
to only a single calorimeter cluster, no other calorimeter clusters,
and requiring that the track precisely point back to the collision
point.  Such a selection is 60\% efficient and has a resolution of 
0.60\% on the momentum of the track in the American L detector.
In the smeared events the resolution on the event energy is extracted
by looking beyond the end point of collision and in this scan is measured
to be $0.803 \pm 0.080$ GeV.  With this resolution the \dlde\ spectrum
and the total
luminosity can be extracted with an error limited by the statistics
of the barrel Bhabha sample and a size essentially given by the
statistics of the Bhabha's.  

	The second simplest region is to consider is the forward region,
for example $0.99824 < |\cos\theta| < 0.99987$ with a cross section
of 33200 pb.  Here forward calorimetry is the only option for measuring
the event energy.  For single electrons making only one calorimeter
cluster the efficiency is 90\% and the energy resolution is 1\%.  In
smeared events the event energy resolution is again extracted with
the beyond the end point technique at is $2.702 \pm 0.018$ GeV.  Again
this resolution can be used to extract the \dlde\ spectrum and total
luminosity to an accuracy given by the statistics of the Bhabha sample.
In this case the error on the luminosity in this scan is 0.05\%, which
is below the theoretical uncertainty claimed for the current generation
of predictions.\cite{bhlumi}

	A feature of the American S detector is forward silicon tracking
disks.	 They occupy the $0.9074 < |\cos\theta| < 0.9903$ region
of the S detector with a cross section of 422 pb, and can be added to any
of the detector designs considered for a high energy $e^+e^-$ linear
collider.  The analysis technique here is to measure the acollinearity
between the two scattered particles and with the nominal beam
energy extract the event energy.\cite{japanese}  The single electron 
analysis requires there to be only one calorimeter cluster and a track
which hits all four planes in the forward silicon tracker.  This gives
a very good resolution on the polar angle of the track of 0.02 milliradians
with an efficiency of 85\%.  Then noting that the
resolution on the event energy is related to the
resolution on the acollinearity by $\sigma_{\sqrt{s}} = \sigma_A p/\sin\theta$,
which in this case gives 18 MeV, it is easy to see that this
technique measures the \dlde\ and the total luminosity only limited by
the statistics of the Bhabha sample.

	Finally in the endcap region of the detector,
$0.829 < |\cos\theta| < 0.996$ with a Bhabha cross section of 1130 pb,
any and all of the analysis techniques described above can be used.  The
tracking resolution on the momentum degrades from the barrel
as the legs of the Bhabha
become more forward but this is usually easy to measure and understand
being dominantly an effect of the decreasing number of hits
on the tracks.  The calorimeter energy resolution can easily be held
to 1\% as in the forward region.  The acollinearity resolution worsens
to 0.03-0.05 milliradians, but this still gives a very good event energy
resolution.  Track based techniques are 60\% efficient and calorimeter
based techniques 90\%.  Here the measurement of \dlde\ and total luminosity
is again limited by the statistics of Bhabha sample and in this
scan gets into the range of the theoretical error on the cross section.

	All these techniques of extracting the \dlde\ spectrum and
the total luminosity from Bhabha scattering at a high energy $e^+e^-$
linear collider were tried on the four different NLC machine parameter sets
given in Table~\ref{tab:beams} and dependence on them is negligible.
Since the extraction is limited by Bhabha statistics whatever machine gives
the highest luminosity is the preferred machine from this point of view.
This also implies that whatever technique samples the largest Bhabha
cross section is preferred.

	A caveat to both of these conclusions is that the simulations
used here do not contain
any treatment of beam related backgrounds.  These have tremendous dependence
on the details of the machine, especially in regards to the amount
of beamstrahlung which besides smearing \dlde\ also produces large numbers
of low energy pairs in the forward direction which can degrade
the resolution in the forward regions of the detector.
To conclude that the best
machine is the highest luminosity machine and that a forward calorimeter
or forward tracker is the best place to measure \dlde\ and the total luminosity
this study needs to be repeated with the effects of beam related background
considered.  Nevertheless it should also be pointed out that a
measurement based on the possible techniques in the
endcaps of the proposed detectors are likely to be insensitive to
the details of the machine induced backgrounds and achieve an accuracy
similar to the current theoretical uncertainty. 


	The effects of the the \dlde\ spectrum on the \ttb\ threshold
shape are quite large.  Figure~2(a) shows the bare $e^+e^- \to t \bar t$
\begin{figure}
\begin{picture}(50,90)(0,0)
\put( 60,0){\epsfig{file=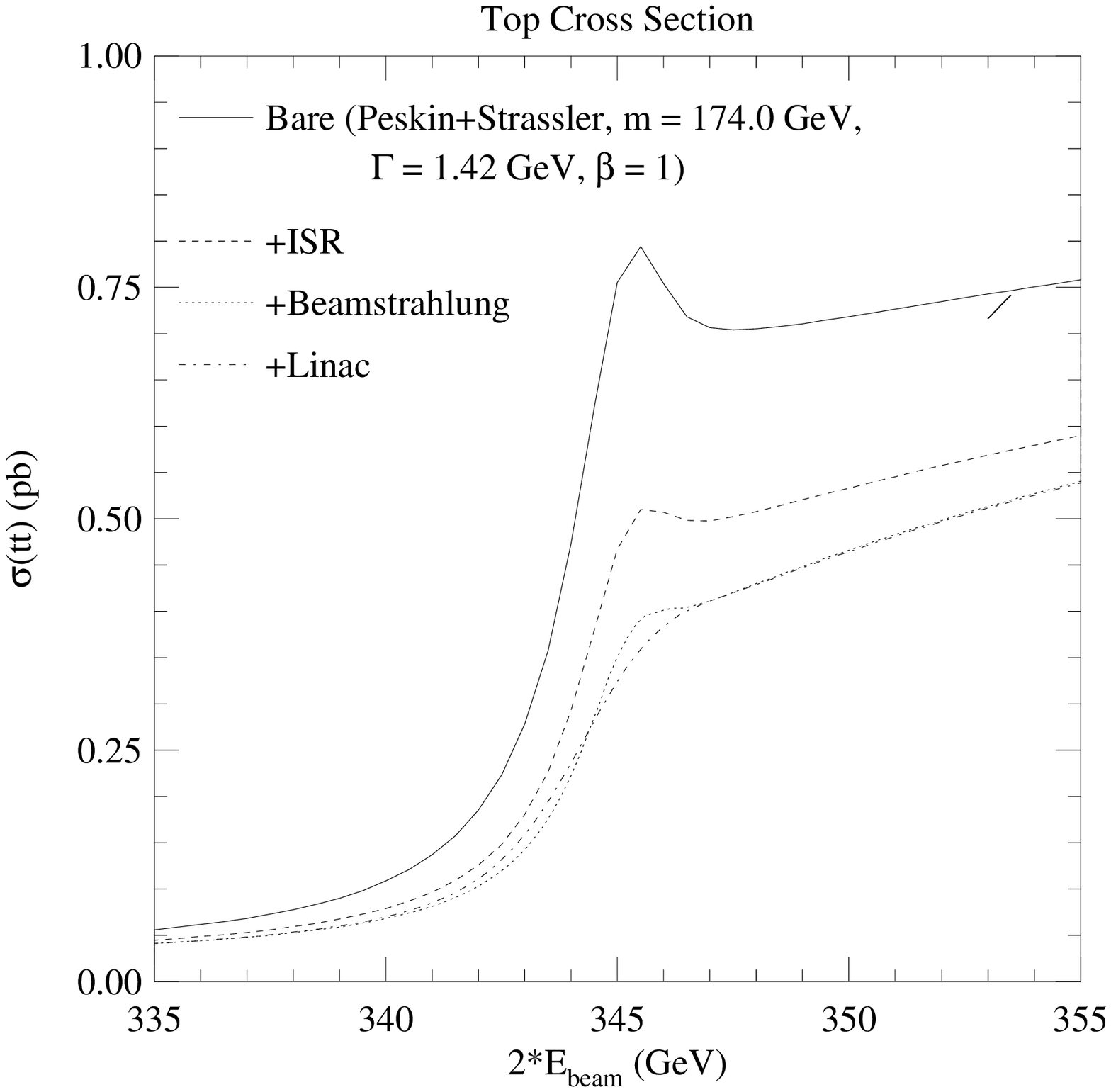,height=1.5in}}
\put (80,50){(a)}
\put(170,0){\epsfig{file=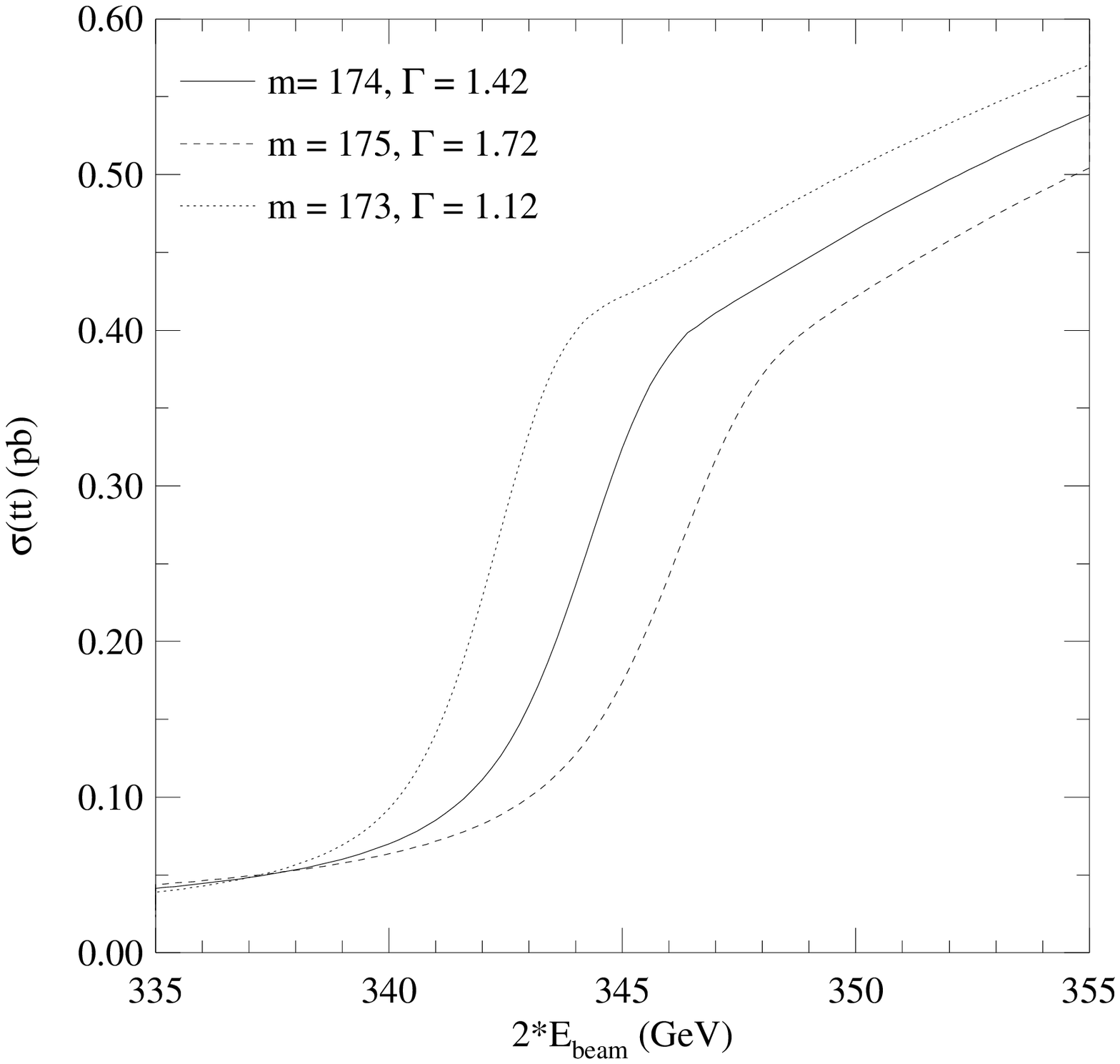,height=1.5in}}
\put(190,50){(b)}
\end{picture}
\caption{(a) The bare $e^+e^- \to t \bar t$ cross section and how it is
smeared by machine effects.  $\beta$ is a measure
of the strength of the top-higgs interaction.  1.0 represents
the prediction of the Standard Model.  (b) The smeared
$e^+e^- \to t \bar t$ cross section for various choices
of top quark parameters.}  
\end{figure}
cross section~\cite{peskin+strasler}
and successively the effects that smear the \dlde\ spectrum have on
that shape.   By the end after ISR, beamstrahlung, and linac energy
spread have been applied there is no sharp peak, but simply a smooth
increase in the cross section.  Figure~2(b) shows the smeared \ttb\ cross
section for different choices of the parameters of the top quark.
This shows that the mass of the top quark effects where the
rise in the cross section will take place and the width will control the
steepness of the rise.

	I can estimate how the measurement of the \dlde\ will effect the
extraction of the parameters of the top quark in a scan of the \ttb\ threshold
region.  The assumptions of this study are simplistic, but seek to isolate
the effects of luminosity spectrum measurement on the extraction of 
the top parameters.  A brief scan is assumed with a total of 2.5/fb 
equally divided into five beam energies: two below the rise; one
at its center; and two at the top.  Unrealistically the number of observed
\ttb\ events is assumed to be measured with an efficiency of one and no
background.  Sources of error considered are the statistics of \ttb\
events and an error on the total luminosity of 0.5\%, conservatively
based on the studies described in above.  Such a scan is repeated
100 times and the resulting cross sections are fit to shapes
with the mass and width as parameters.  The width of the 
distribution of best fit top masses and widths are taken as the errors.

	This study yields a
50~MeV error on the top mass and a 35~MeV on the top width with the
errors being equally contributed from the statistics of \ttb\ events
and the luminosity.  The study is repeated for the four NLC parameter
sets of Table~\ref{tab:beams} and I note that the error on the
width shows significant dependence on the machine parameters.  
For the B parameter set the error on the width is 10\% larger than
the A set.  This is
mainly an effect of the \dlde\ spectrum which is the most smeared for
the B parameters.  This smearing results in fewer \ttb\ events at the
scan points on the top of the rise and thus an increased error on
the steepness of the rise and the top width.  Thus I do conclude
that the parameters of the machine do matter from this point of view.
While higher luminosity is generally preferred it is worth sacrificing
$\sim$10\% of the luminosity for a sharper \dlde\ distribution.



	While at first glance the effect of ISR, beamstrahlung,
and the linac energy spread seem daunting on the possibility of extracting
top quark parameters in a scan of the \ttb\ threshold at a high energy $e^+e^-$
linear collider, this detailed study shows that such extraction
is not limited by the \dlde\ spectrum.  The \dlde\ spectrum
can be measured to an accuracy limited by the statistics of Bhabha
scattering in various techniques using various features of the proposed
detectors.  The many possibilities suggests that the systematic errors
of such a measurement can be limited by extensive cross checks.  Many
other ideas for measuring the \dlde\ spectrum still remain to be explored, and
such explorations are likely to improve the picture.  A key caveat is
the effect of beam induced backgrounds which could severely limit
the use of the forward regions in measuring the \dlde\ spectrum.  This
is a critical area for further study.

	As much as the \ttb\ threshold can be used as the prototype of
a sharp feature in the $e^+e^- \to$ hadrons cross section my studies
do reveal that the optimal high energy $e^+e^-$
linear collider is the one with the highest luminosity.  There
is some gain on the accuracy of the extraction the parameters
of a sharp feature, the width of top quark is an example, by sharpening
the \dlde\ spectrum at the expense of 10\% of the luminosity, but
since the statistics in both the total luminosity and the events
produced by the sharp feature are a major factor in determining the
errors on the feature's parameters a further reduction in the luminosity
to sharpen the \dlde\ spectrum would not be useful. 

\section*{Acknowledgments}

	My research efforts are supported by the US NSF.  I would like
to thank 
Dave Strom, Andreas Kronfeld, David Gerdes,
Charley Baltay, and Dave Burke for useful discussions.

\end{document}